\documentclass[twocolumn,showpacs,preprintnumbers,nofootinbib,amsmath,amssymb]{revtex4}

\usepackage{euscript,amssymb,amsmath}

\usepackage{graphicx}
\usepackage{epsfig}

\newcommand{\be}[1]{\begin{equation}\label{#1}}
\newcommand{\ee}{\end{equation}}
\newcommand{\ba}[1]{\begin{eqnarray}\label{#1}}
\newcommand{\ea}{\end{eqnarray}}
\newcommand{\rf}[1]{(\ref{#1})}
\newcommand{\nn}{\nonumber}

\begin{document}

\title{Kaluza-Klein models: can we construct a viable example?}

\author{Maxim Eingorn}\email{maxim.eingorn@gmail.com}  \author{Alexander Zhuk\ddag}\email{ai_zhuk2@rambler.ru}

\affiliation{Astronomical Observatory and Department of Theoretical Physics, Odessa National University, Street Dvoryanskaya 2, Odessa 65082, Ukraine\\}

\affiliation{\ddag Faculdade de F\'{i}sica, Universidade Federal do Par\'{a}, 66075-110, Bel\'{e}m, PA, Brazil}

\begin{abstract}
In Kaluza-Klein models with toroidal compactification of the extra dimensions, we investigate soliton solutions of Einstein equation. The nonrelativistic gravitational
potential of these solitons exactly coincides with the newtonian one. We obtain the formulas for perihelion shift, deflection of light, time delay of radar echoes and
PPN parameters. Using the constraint on PPN parameter $\gamma$, we find that the solitonic parameter $k$ should be very big: $|k|\geq 2.3\times10^4$. We define a soliton
solution which corresponds to a point-like mass source. In this case the soliton parameter $k=2$, which is clearly contrary to this restriction. Similar problem with the
observations takes place for static spherically symmetric perfect fluid with the dust-like equation of state in all dimensions. The common for both of these models is
the same (dust-like) equations of state in our three dimensions and in the extra dimensions. All dimensions are treated at equal footing. This is the crucial point. To
be in agreement with observations, it is necessary to break the symmetry (in terms of equations of state) between the external/our and internal spaces.

It takes place for black strings which are particular examples of solitons with $k\to \infty$. For such $k$, black strings are in concordance with the observations.
Moreover, we show that they are the only solitons which are at the same level of agreement with the observations as in general relativity. Black strings can be treated
as perfect fluid with dust-like equation of state $p_0=0$ in the external/our space and very specific equation of state $p_1=-(1/2)\varepsilon$ in the internal space.
The latter equation is due to negative tension in the extra dimension. We also demonstrate that dimension 3 for the external space is a special one. Only in this case we
get the latter equation of state. We show that the black string equations of state satisfy the necessary condition of the internal space stabilization. Therefore, black
strings are good candidates for a viable model of astrophysical objects (e.g., Sun) if we can provide a satisfactory explanation of negative tension for particles
constituting these objects.
\end{abstract}


\maketitle

\vspace{.5cm}


\section{\label{sec:1}Introduction}

Any physical theory is only worthy of serious attention when it is consistent with observations. It is well known that general relativity in four-dimensional spacetime
is in good agreement with gravitational experiments such as perihelion shift, deflection of light, time delay of radar echoes and parameterized post-Newtonian (PPN)
parameters. On the other hand, the modern theories of unification such as superstrings, supergravity and M-theory have the most self-consistent formulation in spacetime
with extra dimensions \cite{Polchinski}. Different aspects of the idea of the multidimensionality are intensively used in numerous modern articles. Therefore, it is
important to verify these theories as to their conformity with the experimental data. It was the main aim of our previous paper \cite{EZ3}. We considered a Kaluza-Klein
model with toroidal compactification of the extra dimensions. A matter source was taken in the form of a point-like mass. This approach works very well in general
relativity for calculation in a weak field limit of the formulas for the gravitational experiments \cite{Landau}. We expected that such approach will be also applicable
to our multidimensional model. To verify it, we obtained the metric coefficients in a weak field approximation and applied them to calculate the formulas for
gravitational experiments. We found that expressions for perihelion shift, light deflection, time delay and PPN parameters demonstrate good agreement with the
experimental data only in the case of ordinary three-dimensional space. This result does not depend on the size of the extra dimensions. Therefore, the point-like
gravitational sources are in concordance with experiments only in three-dimensional space. This result was a surprise for us and motivated us to write the present
article to clarify the reason.

The paper is structured as follows. In Section 2 we consider the family of exact 5-D soliton solutions.  To define the connection with our previous paper, among of these
solutions we single out one with asymptotic metric coefficients exactly coinciding with ones obtained in \cite{EZ3} for a point-like mass. Hence, this soliton
contradicts the experiments. We show that $T_{00}$ is the only non-zero component of the energy-momentum tensor in this case. From this point this soliton can be treated
as perfect fluid with the dust-like equation of state in all dimensions.

To demonstrate that the delta-shaped form is not a cause for contradictions with experiments, we consider in Section 3 the finite size static spherically symmetric
perfect fluid with dust-like equation of state in all dimensions. Here, we arrived at exactly the same form of the asymptotic metric coefficients as in the case of
point-like mass source. Therefore, this model also contradicts the observations. Thus, both point-like mass model and perfect fluid with dust-like equation of state in
all directions failed with experiments. The common for both of these models is the same equations of state in our three dimensions and in the extra dimensions. So, all
dimensions are treated at equal footing. This is the crucial point. To be in agreement with observations, it is necessary to break the symmetry between the external/our
and internal spaces.

To prove it, we investigate in Section 4 the conditions under which the solitonic solutions do not contradict the observations. We obtain the formulas for perihelion
shift, deflection of light, time delay of radar echoes and PPN parameters. Using the constraint on PPN parameter $\gamma$ which comes from the Cassini spacecraft
experiment, we found that the solitonic parameter $k$ should be very big: $|k|\geq 2.3\times10^4$. Roughly speaking, $|k|\to \infty$. In the case of point-like mass
soliton solution $k=2$, which is clearly contrary to this restriction.

In Section 5, we consider black strings which are a particular case of the soliton solutions and satisfy the condition $|k|\to \infty$. Four-dimensional part of this
metrics exactly coincides with Schwarzschild metrics and the internal space is flat. Obviously, the results of gravitational experiments in this model exactly coincide
with general relativity. Here, $T_{00}$ and $T_{44}$ are the only non-zero components of the energy-momentum tensor. $T_{44}$ is negative and is called tension.
Moreover, ${T^0}_0=2{T^4}_4$. It can be treated as dust-like equation of state $p_0=0$ in the external space and very specific equation of state $p_1=-(1/2)\varepsilon$
in the internal space. Additionally, we consider in this section the static spherically symmetric perfect fluid with dust-like equation of state $p_0=0$ in
$d_0$-dimensional external space and an arbitrary equation of state $p_1=\omega \varepsilon$ in $d_1$-dimensional internal space. We demonstrate that the demand of an
exact correspondence between this model and general relativity automatically leads to equation of state $p_1=-(1/2)\varepsilon$ for $d_0=3$. The dimension 3 for the
external space is a special one. Only in this case parameter $\omega $ does not depend on $d_1$ and equals $-1/2$. Therefore, in the case of three-dimensional external
space, the black string equations of state $p_0=0$ and $p_1=-(1/2)\varepsilon$ are the only ones which ensure the same level of agreement with the observations as in
general relativity.

The main results are summarized and discussed in the concluding Section 6. Here, we reveal one very important property of black strings. We show that the black string
equations of state satisfy the necessary condition of the internal space stabilization. Therefore, black strings are good candidates for a viable model of astrophysical
objects (e.g., Sun) if we can provide a satisfactory explanation of negative tension for particles constituting these objects.




\section{\label{sec:2}Soliton metrics}

The point-like matter source is the reasonable approximation in 3-dimensional space in the case when the distance to a gravitating mass is much greater than its radius.
This approximation was used, e.g., in \cite{Landau} to get formulas for perihelion shift and deflection of light in general relativity. At first glance, this approach
should also work well in the case of a multidimensional space. To check this assumption, in our paper \cite{EZ3} we obtained asymptotic expression for the metric
coefficients in multidimensional spacetime with the point-like mass $m$ at rest:
\ba{1.0} &{}&ds^2\approx\left(1-\frac{r_g}{r_3}+\frac{r_g^2}{2r_3^2} \right)c^2dt^2\nn \\
&-&\left(1+\frac{1}{D-2}\, \frac{r_g}{r_3}\right)\left(dr_3^2+r_3^2d\theta^2+r_3^2\sin^2\theta d\psi^2\right)\nn \\
&-&\left(1+\frac{1}{D-2}\, \frac{r_g}{r_3}\right)\left((dx^4)^2+(dx^5)^2+\ldots + (dx^D)^2\right)\, , \nn \\
\ea
where $r_3$ is the length of a radius vector in three-dimensional space, $r_g=2G_Nm/c^2$, $G_N$ is the newtonian gravitational constant and we used three-dimensional
isotropic coordinates. We suppose that the $(D=3+d)$-dimensional space has the factorizable geometry of a product manifold $M_D=\mathbb{R}^3\times T^{d}$. $\mathbb{R}^3$
describes the three-dimensional asymptotically flat external (our) space and $T^{d}$ is a torus which corresponds to a $d$-dimensional internal space with volume $V_d$.
Then, we demonstrated that this metrics does not provide the correct values of the classical gravitational tests (perihelion shift, light deflection, PPN parameters) for
$D>3$. Mathematically, this discrepancy arises because of the prefactor $1/(D-2)$ in metric coefficients  \rf{1.0} instead of the prefactor 1 as in general relativity.

On the other hand, there is a number of well known exact vacuum solutions for the Kaluza-Klein models. Therefore,  it is of interest to determine the relationship
between these exact solutions and our asymptotic metric coefficients and try to understand why the delta-shaped matter source approach does not work in multidimensional
space. In this regard, we will investigate 5-D static metrics in isotropic (with respect to our three-dimensional space) coordinates:
\be{1.1} ds^2=A(r_3)c^2dt^2+B(r_3)\left(dx^2+dy^2+dz^2\right)+C(r_3)d\xi^2\, , \ee
where $r_3=\sqrt{x^2+y^2+z^2}$. This spacetime has two Killing vectors $\partial/\partial t$ and $\partial/\partial \xi$. It is clear that appropriate energy-momentum
tensor also should not depend on time $t$ and fifth coordinate $\xi$. We suppose that metrics \rf{1.1} is a solution of the vacuum Einstein equation
\be{1.2} R_{ik}=0 \ee
with the proper boundary conditions. It is worth of noting that the dependence of the metric coefficients in \rf{1.1} only on $r_3$ means that the matter source for such
metrics is uniformly "smeared" over the fifth dimension \cite{EZ1,EZ2}. It is clear that in this case the non-relativistic gravitational potential depends only on $r_3$
and exactly coincides with the newtonian one.

To our knowledge, the first solution of the form of \rf{1.1} in non-isotropic "Schwarzschild-like" coordinates was found in \cite{Kramer} and reads
\ba{1.3} &{}&ds^2=\left(1-\frac{b}{r_3'}\right)^{a'}c^2dt^2
-\left(1-\frac{b}{r_3'}\right)^{-a'-b'}dr_3'^2\nn \\
&-&\left(1-\frac{b}{r_3'}\right)^{1-a'-b'}r_3'^2d\Omega_2^2-\left(1-\frac{b}{r_3'}\right)^{b'}d\xi^2\, , \ea
where $a'$ and $b'$ are constants satisfying the condition
\be{1.4} a'^2+a'b'+b'^2=1\, \ee
and the parameter $b$ is usually connected with the gravitating mass: $a'b=2G_Nm/c^2 =r_g$. Then, in the isotropic coordinates this solution was obtained in
\cite{soliton} and \cite{Davidson} and dubbed in the literature the soliton solution. Its generalization for $D\ge 5$ was performed in \cite{Ivashchuk,FIM,Leon}. In our
paper we choose the metrics in the parametrization proposed in \cite{Davidson}:
\ba{1.5} &{}&ds^2=\left(\frac{ar_3-1}{ar_3+1}\right)^{2\varepsilon k}c^2dt^2\nn \\
&-&\left(1-\frac{1}{a^2r_3^2}\right)^2\left(\frac{ar_3+1}{ar_3-1}\right)^{2\varepsilon(k-1)}\left(dr_3^2+r_3^2d\Omega_2^2\right)\nn \\
&-&\left(\frac{ar_3+1}{ar_3-1}\right)^{2\varepsilon}d\xi^2\, , \ea
where $a,\varepsilon$ and $k$  are constants  and parameters $\varepsilon$ and $k$ satisfy the condition
\be{1.6} \varepsilon^2\left(k^2-k+1\right)=1\, . \ee
The Schwarzschild-like solution \rf{1.3} and the soliton solution \rf{1.5} are connected by the relations:
\be{1.7} r_3'=r_3\left(1+\frac{b}{4r_3}\right)^2 \ee
and
\be{1.8} a'=\varepsilon k,\ b'=-\varepsilon,\ a=\frac{4}{b}\, . \ee
It follows from \rf{1.7} that $r_3'=r_3 +O(1/c^2)$ if $b=4/a=r_g/a'$.

In the approximation $f\equiv 1/(ar_3) \ll 1$ and up to $O(f)$ we obtain for the metric coefficients of \rf{1.5} the following formulas:
\ba{1.9} &{}&B(r_3)=-\left(1-f^2\right)^2\left(\frac{1+f}{1-f}\right)^{2\varepsilon(k-1)}\nn \\
&\approx&-1-4\varepsilon(k-1)f=-1-\frac{4\varepsilon(k-1)}{ar_3} \ea
and
\be{1.10} C(r_3)=-\left(\frac{1+f}{1-f}\right)^{2\varepsilon}\approx -1-4\varepsilon f=-1-\frac{4\varepsilon}{ar_3}\, . \ee
Now, we want to compare these expressions with asymptotic metric coefficients from \rf{1.0} (where $D=4$):
\be{1.11} B(r_3)\approx-1-\frac{r_g}{2r_3},\ \ \ C(r_3)\approx-1-\frac{r_g}{2r_3}\, . \ee
Comparing expressions \rf{1.9} and \rf{1.10} with our asymptotic metric coefficients \rf{1.11}, we get
\be{1.12} k=2,\ \varepsilon=\frac{1}{\sqrt{3}},\ a=\frac{8}{\sqrt{3}r_g}\, , \ee
where we take into account the relation \rf{1.6}. Finally, for $A(r_3)$ from \rf{1.5} up to $O(f^2)$ we get
\ba{1.13}
&{}&A(r_3)=\left(\frac{1-f}{1+f}\right)^{2\varepsilon k}\approx1-4\varepsilon kf+8\varepsilon^2k^2f^2\nn \\
&=&1-\frac{4\varepsilon k}{ar}+\frac{8\varepsilon^2k^2}{a^2r^2}=1-\frac{r_g}{r}+\frac{r_g^2}{2r^2} \ea
in complete analogy with asymptotic metric coefficient $A(r_3)$ in \rf{1.0}. Therefore, for the parameters \rf{1.12} the soliton metrics \rf{1.5} reads
\ba{1.14} &{}&ds^2=\left(\frac{1-\sqrt{3}r_g/8r_3}{1+\sqrt{3}r_g/8r_3}\right)^{\frac4{\sqrt{3}}}c^2dt^2\nn \\
&-&\left(1-3r_g^2/64r_3^2\right)^2\left(\frac{1+\sqrt{3}r_g/8r_3}{1-\sqrt{3}r_g/8r_3}\right)^{\frac2{\sqrt{3}}}
\left(dr_3^2+r_3^2d\Omega_2^2\right)\nn \\
&-&\left(\frac{1+\sqrt{3}r_g/8r_3}{1-\sqrt{3}r_g/8r_3}\right)^{\frac2{\sqrt{3}}}d\xi^2\, . \ea
Our analysis shows that this form of metrics provides the correct asymptotic behavior in the case of delta-shaped matter source. The metrics \rf{1.14} is the exact
solution of the Einstein equation for the gravitating mass at rest (${\bf v}\equiv 0$) uniformly "smeared"  over the extra dimension. The only non-zero component of the
energy-momentum tensor is $T_{00}$.
We can prove it by the following way. It is clear from the previous consideration that the metric coefficients in \rf{1.14} up to the terms $1/c^2$ read
\ba{1.15} &{}&g_{00}\approx1-\frac{r_g}{r_3}=1+h_{00}\, ,\nn \\
&{}&g_{\alpha\alpha}\approx-1-\frac{r_g}{2r_3}=
-1+h_{\alpha\alpha}\quad\Rightarrow\nn \\
&{}&h_{00}=-\frac{r_g}{r_3}\, , \quad h_{\alpha\alpha}=-\frac{r_g}{2r_3}\, , \quad  \alpha =1,2,3,4\, . \ea
With the same accuracy  the components of the Ricci tensor are
\ba{1.16} &{}&R_{00}\approx\frac{1}{2}\triangle h_{00}=-\frac{1}{2}\triangle\frac{r_g}{r_3}=-\frac{G_Nm}{c^2}\triangle\frac{1}{r_3}\nn \\
&=&\frac{4\pi G_Nm\delta({\bf r}_3)}{c^2}=k_N\,\frac{1}{2}m\delta({\bf r}_3)c^2\, ,\nn \\
&{}&R_{\alpha\alpha}\approx k_N\, \frac{1}{4}m\delta({\bf r}_3)c^2\, , \quad  \alpha =1,2,3,4\, , \ea
where $k_N\equiv 8\pi G_N/c^4$ and
$\triangle = \delta^{\alpha\beta}\partial^2/\partial x^{\alpha}\partial x^{\beta}$ is the $D$-dimensional Laplace operator (see for details \cite{EZ3}). \rf{1.16}
indicate that all spatial components of the Ricci tensor are equal to each other. From Einstein equations easily follows that this equality may take place only if
$T_{11}=T_{22}=T_{33}=T_{44}$. Taking into account that the matter source is at rest: $T_{11}=T_{22}=T_{33}= 0\quad \Rightarrow \quad T_{44} =0$, we may conclude
that the only non-zero component of the energy-momentum tensor is $T_{00}$. Hence, for the Einstein equations we obtain
\ba{1.17} R_{00}&=&k_{\mathcal{D}}\left(T_{00}-\frac{1}{3}Tg_{00}\right) = k_{\mathcal{D}}\, \frac{2}{3}T_{00},\nn \\
R_{\alpha\alpha} &=& k_{\mathcal{D}}\left(T_{\alpha\alpha}-\frac{1}{3}Tg_{\alpha\alpha}\right)\approx k_{\mathcal{D}}\, \frac{1}{3}T_{00}\, , \quad  \alpha =1,2,3,4\, ,\nn\\ \ea
where $k_{\mathcal{D}}\equiv 2S_D\tilde G_{\mathcal{D}}/c^4$ and $S_D=2\pi^{D/2}/\Gamma (D/2)$ is the total solid angle (surface area of the $(D-1)$-dimensional sphere of a unit
radius), $\tilde G_{\mathcal{D}}$ is the gravitational constant in the $(\mathcal{D}=D+1)$-dimensional spacetime.

Therefore, from \rf{1.16} and \rf{1.17} up to the terms $c^2$ we get
\be{1.18} T_{00}\approx\frac{k_N}{k_{\mathcal{D}}}\, \frac{3}{4}m\, \delta({\bf r}_3)c^2 = \frac{1}{a_1}\, m\, \delta({\bf r}_3)c^2\, , \ee
where $a_1$ is the size of the extra dimension (the period of the torus) and we take into account the relation between the gravitational constants \cite{EZ2}:
\be{1.19} \frac{2(D-2)}{(D-1)}S_D \, \tilde G_{\mathcal{D}}/\prod_{\alpha=1}^d a_{\alpha} =4\pi G_N \, . \ee
We can write \rf{1.18} in the form $T_{00}\approx \varepsilon \approx \rho c^2$ where $\rho \equiv (m/a_1)\, \delta({\bf r}_3)$ is the rest mass density in the case of a
point mass smeared over the extra dimension. Thus, for the considered model the energy-momentum tensor reads ${T^i}_k={\mbox{diag}}\left(\varepsilon , 0,0,0,0\right)$
and corresponds to the delta-shaped matter source. Unfortunately, despite this clear physical interpretation, solution \rf{1.14} contradicts the observations. As we have
mentioned above, the relation $h_{\alpha\alpha}=(1/2)h_{00}$ (see \rf{1.15}) results in formulas for the classical gravitational tests which predict effects considerably
different from the observations \cite{EZ3}. It seems that the reason for this is delta-shaped form of the source. However, it is not. To show it, in the next section we
consider a finite size spherically symmetric perfect fluid.

\section{\label{sec:3}Static spherically symmetric perfect fluid with dust-like equation of state in all dimensions}

Let us consider now perfect fluid which fills uniformly a three-dimensional sphere of the radius $R$ and one-dimensional internal torus with the period $a_1$. If $m$ is
a non-relativistic rest mass of this configuration, then rest mass density reads
\be{2.1} \rho =\frac{m}{a_1V_3}\equiv \frac{\rho_3}{a_1}\; ,\quad {\mbox{where}}\quad \rho_3=\frac{m}{(4/3)\pi R^3}\, . \ee
The energy-momentum tensor is taken in the form
\be{2.2} {T^i}_k = \left\{\begin{array}{ccc}
&{\mbox{diag}}\left(\varepsilon,-p_0,-p_0,-p_0,-p_1\right)\, ;& r_3\leq R\, , \forall \xi\\
&0\; ;& r_3>R\, , \forall \xi
\end{array}\right.
\ee
where $\xi \in [0, a_1]$ is a coordinate of the extra dimension. We shall consider the case when the energy density is much bigger than the pressure in all dimensions:
\be{2.3} \varepsilon \gg p_0,p_1\, . \ee
This is the usual condition for astrophysical objects in three-dimensional space.

Now, we want to calculate the metric coefficients $g_{ik}\quad (i,k =0,1,2,3,4)$ in the weak field limit. Prior to that, we would like to make two comments. First, it is
well known that for static configurations the non-diagonal metric coefficients are absent: $g_{0\alpha}=0\quad (\alpha =1,2,3,4)$. Second, because of spherical symmetry
in three-dimensional space and uniform distribution of matter over the internal space, the metric coefficients depend only on $r_3$: $\; g_{ik}=g_{ik}(r_3)$. In the weak
field limit the metric coefficients take the form:
\be{2.4} g_{00}\approx 1+h_{00}\, ,\quad g_{\alpha\alpha}\approx -1+h_{\alpha\alpha}\, , \quad h_{00},h_{\alpha\alpha} \sim O\left(1/c^2\right)\, . \ee
In particular,  $h_{00}\equiv 2\varphi/c^2$. Later we will demonstrate that $\varphi$ is the non-relativistic gravitational potential. It can be easily shown (see, e.g.,
\cite{EZ3}) that Ricci tensors up to the order $1/c^2$ read
\be{2.5} R_{00}\approx \frac12 \triangle h_{00}\, ,\quad  R_{\alpha\alpha}\approx \frac12 \triangle h_{\alpha\alpha}\, , \ee
where $D$-dimensional Laplace operator $\triangle $
is reduced to usual three-dimensional one. The conditions $T_{00}\gg T_{0\alpha},T_{\alpha\beta}$ result in Einstein equations of the form of \rf{1.17} (the only
difference consists in replacement of the exact equality in the first equation by the approximate one). Therefore, we arrive at the following relation between the metric
coefficients:
\be{2.6} h_{\alpha\alpha}=\frac12 h_{00}=\frac{\varphi}{c^2}\, . \ee
We can easily see that the Einstein equations in this approximation are reduced to the Poisson equation for the non-relativistic gravitational potential $\varphi$:
\ba{2.7}
\triangle \varphi^{(in)} &=&4\pi G_N\rho_3\, ,\nn \\
\triangle \varphi^{(out)}&=&0\, , \ea
where we took into account that in the inner region ($r_3\leq R$) $\; T_{00}^{(in)}\approx \varepsilon \approx \rho c^2 =\rho_3c^2/a_1$ and in the outer region ($r_3 >
R$) $\; T_{00}^{(out)}=0$. The latter equation in \rf{2.7} has the solution
\be{2.8} \varphi^{(out)} =-\frac{G_N m}{r_3}\, , \ee
where we used the physical boundary condition $\varphi^{(out)}\to 0 $ for $r_3\to +\infty$. Therefore, relation \rf{2.6} leads to the same expressions for
$h_{00}^{(out)}$ and $h_{\alpha\alpha}^{(out)}$ as in \rf{1.15}. It means that considered perfect fluid model contradicts the observations.

Thus, both point-like mass model and perfect fluid with dust-like equation of state in all directions failed with experiments. At first glance, it is very strange result
which works against KK models because such matter sources have clear physical motivation. However, as we shall see below, such approach is naive. The common for both of
these models is the same equations of state in our three dimensions and in the extra dimensions. In considered models, it is dust-like equation of state. So, all
dimensions are treated at equal footing. This is the crucial point. As we shall see, to satisfy the observations we should break this symmetry (in terms of equations of
state) between our and extra dimensions.

\section{\label{sec:4}Experimental restrictions on solitons}

Now, we want to demonstrate that experimental restrictions on parameters of soliton solutions \rf{1.3} or \rf{1.5} result in breaking of symmetry between our and extra
dimensions. To show this, we consider parameterized post-Newtonian (PPN) parameters for considered metrics.

It is well known (see, e.g., \cite{Will,Straumann}) that in PPN formalism the static spherically symmetric metrics in isotropic coordinates reads
\be{3.1} ds^2=\left(1-\frac{r_g}{r_3}+\beta\frac{r_g^2}{2r_3^2} \right)c^2dt^2 -\left(1+\gamma \frac{r_g}{r_3}\right)\sum_{i=1}^{3}\left(dx^i\right)^2 \, . \ee
In general relativity $\beta=\gamma=1$. To get $\beta$ and $\gamma$ in the case of soliton solution \rf{1.5}, it is sufficient  to analyze the asymptotic expressions
\rf{1.9} and \rf{1.13} for the metric coefficients $B(r_3)$ and $A(r_3)$, respectively. As we already mentioned above, in the case of smeared extra dimension the
non-relativistic gravitational potential should exactly coincide with the Newtonian one. Because the function $A(r_3)$ is the metric coefficient $g_{00}$, this demand
leads to the following condition:
\be{3.2} \frac{4\varepsilon k}{a}=r_g=\frac{2G_Nm}{c^2}\, . \ee
Then, \rf{1.13} reads
\be{3.3} A(r_3) = \left(\frac{1-f}{1+f}\right)^{2\varepsilon k}\approx1-\frac{r_g}{r_3}+\frac{r_g^2}{2r_3^2}\, . \ee
Simple comparison of \rf{3.3} with the metric coefficient $g_{00}$ in \rf{3.1} shows that soliton PPN parameter
\be{3.4} \beta_s=1\, , \ee
as in general relativity. To obtain soliton PPN parameter $\gamma_s$, we need to analyze the asymptotic expression \rf{1.9} for $B(r_3)$. Substitution of the relation
\rf{3.2} into \rf{1.9} gives
\ba{3.5} &{}&B(r_3)=-\left(1-f^2\right)^2\left(\frac{1+f}{1-f}\right)^{2\varepsilon(k-1)}\nn \\
&\approx&-\left(1+\frac{k-1}{k}\frac{r_g}{r_3}\right) \ea
and comparison of equations \rf{3.5} and \rf{3.1} gives
\be{3.6} \gamma_{s}=\frac{k-1}{k}\, . \ee
Now, with the help of these PPN parameters, we can easily get formulas for famous gravitational experiments \cite{EZ3,Will,Will2}:

\vspace{0.4cm}

{\it perihelion shift}
\ba{3.7}
&{}&\delta\psi=\frac{6\pi mG_N}{a\left(1-e^2\right)c^2}\frac{1}{3}(2+2\gamma_s-\beta_s)\nn \\
&=&\frac{6\pi mG_N}{a\left(1-e^2\right)c^2}\frac{3k-2}{3k}=\frac{(3k-2)\pi r_g}{ka\left(1-e^2\right)}\, , \ea

\vspace{0.4cm}

{\it deflection of light}
\be{3.8} \delta\psi=(1+\gamma_{s})\frac{r_g}{\rho}=\frac{2k-1}{k}\frac{r_g}{\rho}\, , \ee

\vspace{0.3cm}

{\it time delay of radar echoes (the Shapiro time delay effect)}
\ba{3.9}
&{}&\delta t=(1+\gamma_{s})\frac{r_g}{c}\ln\left(\frac{4r_{Earth}r_{planet}}{R_{Sun}^2}\right)\nn \\
&=& \frac{2k-1}{k}\frac{r_g}{c}\ln\left(\frac{4r_{Earth}r_{planet}}{R_{Sun}^2}\right)\, . \ea
In \rf{3.7}, the semi-major axis $a$ of the orbit of a planet should not be confused with parameter $a$ of the solution \rf{1.5}.

Comparison of the formulas \rf{3.7}-\rf{3.9} with experimental data gives the possibility to restrict parameters of soliton solutions. However, we can get it directly
from experimental restriction on PPN parameter $\gamma$. The tightest constraint on $\gamma$ comes from  the Shapiro time-delay experiment using the Cassini spacecraft:
$\gamma-1 =(2.1\pm 2.3)\times 10^{-5}$ \cite{Will2,JKh,Bertotti}. Thus, from \rf{3.6} we find that solitonic parameter $k$ should satisfy the condition
$|k|\geq2.3\times10^4$.

Obviously, the point-like mass soliton \rf{1.14} with $k=2$ does not satisfy this restriction. We have shown that $T_{00}$ is the only non-zero component of the
energy-momentum tensor for this solution. It means that we have the same dust-like equation of state in all $D=4$ dimensions. It is not difficult to show that all other
solitons with parameters different from \rf{1.12} besides non-zero $T_{00}$ will also have some other non-zero components of the energy-momentum tensor which destroy the
symmetry between our and extra dimensions. We shall demonstrate it for one particular case of solitons called black strings. It is a unique case because only this
solitonic solution does not prevent the stabilization of the internal space (see the following discussion).

\section{\label{sec:5}Black strings}

Now, we consider a particular example which satisfies the condition $|k|\geq2.3\times10^4$. It corresponds in \rf{1.5} to the limit
\be{4.1} \varepsilon\rightarrow0,\ k\rightarrow+\infty,\ \varepsilon k\rightarrow1\, , \ee
or to the limit $a'\rightarrow1,\ b'\rightarrow0$ in the Schwarzschild-like metrics \rf{1.3}. In this limit the metrics \rf{1.5} reads
\ba{4.2} &{}&ds^2=\left(\frac{ar_3-1}{ar_3+1}\right)^{2}c^2dt^2\nn \\
&-&\left(\frac{ar_3+1}{ar_3}\right)^{4}\left(dr_3^2+r_3^2d\Omega_2^2\right)-d\xi^2\, . \ea
It can be easily seen that the four-dimensional part of this metrics (which corresponds to the section $\xi =\mathrm{const}$) is the pure Schwarzschild metrics 
(for $a= 4/r_g$) in isotropic coordinates. Metrics of the form \rf{4.2} is often called the uniform black string. From this metrics up to the terms $1/c^2$ we get
\ba{4.3}
g_{00}&\approx& 1-\frac{r_g}{r_3}=1+h_{00},\nn \\
g_{\alpha\alpha}&\approx& -1-\frac{r_g}{r_3}=-1+h_{\alpha\alpha}\, , \quad \alpha=1,2,3,\nn \\
g_{44}&=&-1=-1+h_{44}\quad\Rightarrow\nn \\
h_{00}&=&h_{11}=h_{22}=h_{33}=-\frac{r_g}{r_3},\ \ \ h_{44}=0\, . \ea
With the same accuracy  the components of the Ricci tensor are
\ba{4.4}
&{}&R_{\alpha\alpha}\approx\frac{1}{2}\triangle h_{\alpha\alpha}=-\frac{1}{2}\triangle\frac{r_g}{r_3}=-\frac{G_Nm}{c^2}\triangle\frac{1}{r_3}\nn \\
&=&\frac{4\pi G_Nm\delta({\bf
r}_3)}{c^2}=k_N\, \frac{1}{2}m\delta({\bf r}_3)c^2\, ,\quad \alpha =0,1,2,3\, , \nn\\
&{}&R_{44}=0\, . \ea
For a gravitating mass at rest and based on the form of \rf{4.4}, we arrive at the conclusion that $T_{00}$ and $T_{44}$ are the only non-zero components of the
energy-momentum tensor. Thus, the Einstein equations take the form:
\ba{4.5}
R_{00}&=&k_{\mathcal{D}}\left(T_{00}-\frac{1}{3}Tg_{00}\right)\, ,\nn \\
R_{11}&=&R_{22}=R_{33}=k_{\mathcal{D}}\left(-\frac{1}{3}Tg_{11}\right)\, ,\nn \\
R_{44}&=&k_{\mathcal{D}}\left(T_{44}-\frac{1}{3}Tg_{44}\right)\, . \ea
Therefore, from \rf{4.4} and \rf{4.5} up to the terms $c^2$ we get
\ba{4.6} &{}&T_{00}\approx\frac{k_N}{k_{\mathcal{D}}}m\delta({\bf r}_3)c^2 = \frac{4}{3}\frac{m}{a_1}\delta({\bf r}_3)c^2\, ,\nn\\
&{}&T_{44}\approx-\frac{1}{2}\frac{k_N}{k_{\mathcal{D}}}m\delta({\bf r}_3)c^2 =-\frac{2}{3}\frac{m}{a_1}\delta({\bf r}_3)c^2\, ,\nn\\
&{}&T=T_{00}g^{00}+T_{44}g^{44}\nn\\
&\approx&\frac{3}{2}\frac{k_N}{k_{\mathcal{D}}}m\delta({\bf r}_3)c^2=2\frac{m}{a_1}\delta({\bf r}_3)c^2\, . \ea
Non-zero component $T_{44}$ results in non-vanishing tension of the black strings (see, e.g., \cite{Harmark,Traschen}). \rf{4.6} shows that the value $T_{00}$ is in two
times bigger than the absolute value of $T_{44}$. Similar relation exists for the ADM mass and the tension (see e.g. (B.11) and (B.12) in \cite{Harmark}). The numerical
prefactors $4/3$ and $-2/3$ in front of $T_{00}$ and $T_{44}$ follow from the normalization \rf{1.19}. We can choose different relation between $\tilde G_{\mathcal{D}}$
and $G_N$ which will lead to other numerical prefactors. However, the relation $T_{00}\approx-2T_{44}$ will remain the same. If we introduce the energy density
$\varepsilon={T^0}_0\approx T_{00}$, pressure in our external three-dimensional space $p_0=-{T^{\alpha}}_{\alpha} \approx T_{\alpha\alpha}\; (\alpha =1,2,3)$ and
pressure in the internal space $p_1=-{T^4}_4\approx T_{44}$, then the black string energy-momentum tensor can be written in the form
\be{4.7} {T^i}_k ={\mbox{diag}}\left(\varepsilon,-p_0,-p_0,-p_0,-p_1\right) \ee
with equations of state:
\be{4.8} p_0=0 \quad {\mbox{and}}\quad p_1=-\frac12 \varepsilon\, . \ee
Therefore, for black strings there is no symmetry between our and extra dimensions. Additionally, \rf{4.3} show that the relations between the metric coefficients
$h_{00}, h_{11},h_{22}$ and $h_{33}$ are exactly the same as in general relativity and $h_{44}=0$. Hence, in the case of black strings there is no deviations from
formulas of general relativity for gravitational experiments. Moreover, the condition $h_{00}=h_{\alpha\alpha} \; (\alpha =1,2,3)$ automatically leads to equation of
state $p_1=-(1/2)\varepsilon$ and equality $h_{44}=0$ for static spherically symmetric perfect fluid with dust-like equation of state in our space. We shall prove it
now.

Let us consider a static spherically symmetric perfect fluid with energy-momentum tensor
\be{4.9} {T^i}_k = {\rm diag\ } (\, \varepsilon,
\underbrace{-p_0,\ldots ,-p_0}_{\mbox{$d_0$ times}%
}, \; \, \underbrace{-p_1,\ldots ,-p_1}_{%
\mbox{$d_1$ times}}\, )\, . \ee
In what follows, we shall use the notations: $i,k = 0,1,\ldots,D\, ; \; a,b = 1,\ldots ,D\, ; \; \alpha,\beta =1,\ldots,d_0$ and $\mu,\nu=d_0+1,\ldots,d_0+d_1$. For
static spherically symmetric configurations holds $g_{0a}=0$ and $g_{ab}=0\, ,\,  a\neq b$. Because we want to apply this model to ordinary astrophysical objects, we
suppose the dust-like equation of state in $d_0$-dimensional external space: $p_0\approx 0$, but the equation of state is an arbitrary one in $d_1$-dimensional internal
space: $p_1=\omega \varepsilon$.
We still consider the weak field approximation where the metric coefficients can be expressed in the form of \rf{2.4} (where $\alpha$ should be replaced by $a$).
Additional requirement that we impose is that considered configuration does not contradict the observations. Obviously, this will occur if the following conditions hold:
$h_{00}=h_{\alpha\alpha}$ and $h_{\mu\mu}=0$. We now define when it takes place.

First, taking into account that $T=\sum_{i=0}^D {T^i}_i=\varepsilon (1-\omega d_1)\, ,\; T_{\alpha\alpha}=0\, ,\; \varepsilon \sim O(c^2)$ and up to terms $c^2$ that
$T_{00}\approx {T^0}_0\, ,\; T_{\mu\mu}\approx -{T^{\mu}}_{\mu}$, we get from the Einstein equation
\be{4.10} R_{ik} =k_{\mathcal{D}}\left(T_{ik}-\frac{1}{D-1}T g_{ik}\right) \ee
non-zero components of Ricci tensor (up to $1/c^2$):
\ba{4.11}
R_{00}&\approx& \varepsilon k_{\mathcal{D}}\; \frac{d_0-2+d_1(1+\omega)}{D-1}\, ,\\
R_{\alpha\alpha} &\approx& \varepsilon k_{\mathcal{D}}\; \frac{1-\omega d_1}{D-1}\, ,\label{4.12}\\
R_{\mu\mu} &\approx& \varepsilon k_{\mathcal{D}}\; \frac{\omega(d_0-1)+1}{D-1} \label{4.13}\, , \ea
where $k_{\mathcal{D}}\sim O(1/c^4)$ is defined in \rf{1.17}.
On the other hand, the components of Ricci tensor read
\be{4.14} R_{00}\approx \frac12 \triangle h_{00}\, ,\quad  R_{aa}\approx \frac12 \triangle h_{aa}\, , \ee
where as usual we put $h_{00}\equiv 2\varphi/c^2$ and $\triangle$ is $D$-dimensional Laplace operator defined in \rf{1.16}. Therefore, from equations \rf{4.11},
\rf{4.12} and \rf{4.14} we obtain
\be{4.15} h_{\alpha\alpha}=\frac{1-wd_1}{d_0-2+d_1(1+\omega)}h_{00}\, . \ee
As we mentioned above, to be in agreement with experiment we should demand
\be{4.16} h_{\alpha\alpha}=h_{00}\quad \Longrightarrow \quad \omega =\frac{3-D}{2d_1}=\left.-\frac12\right|_{d_0=3}\, . \ee
Hence, for considered perfect fluid the condition $h_{\alpha\alpha}=h_{00}$ results in the following conclusions:

\vspace{0.3cm}

1. {\it The number of dimensions of the external space $d_0=3$ is a special case. Only in this case the parameter of equation of state $\omega$  does not depend on
number of the extra dimensions $d_1$.}

2. {\it For $d_0=3$, the equation of state in the internal space is $p_1=-(1/2)\varepsilon$ which exactly coincides with the case of black strings (see \rf{4.8}).}

\vspace{0.3cm}

Substitution of $\omega$ from \rf{4.16} into \rf{4.13} gives
\ba{4.17} &{}&R_{\mu\mu}\approx\varepsilon k_{\mathcal{D}}\; \frac{(3-d_0-d_1)(d_0-1)+2d_1}{2d_1(D-1)}=\left.0\right|_{d_0=3} \nn\\
&{}&\Longrightarrow\quad h_{\mu\mu}=0|_{d_0=3}\, . \ea
Therefore, as in the case of black strings, we obtain flat extra dimensions.

To conclude the consideration of this perfect fluid, we want to get the metric coefficients. To do it, it is sufficient to define the function $\varphi$. It can be
easily seen from \rf{4.11} and \rf{4.14} that this function satisfies the equation
\be{4.18} \frac{1}{c^2}\, \triangle \varphi = \varepsilon k_{\mathcal{D}}\; \frac{D-2+\omega d_1}{D-1}\, , \ee
where $\varepsilon \approx \rho^{(D)}c^2$ and $\rho^{(D)}$ is $D$-dimensional non-relativistic rest mass density. We want to reduce this equation to ordinary Poisson
equation. We consider the case when matter is uniformly smeared over the extra dimensions: $\rho^{(D)}=\rho^{(d_0)}/V_{d_1}$ where $V_{d_1}$ is the volume of the
internal space. In this case the non-relativistic potential $\varphi$ depends only on our external coordinates and $\triangle$ is reduced to $d_0$-dimensional Laplace
operator. If we demand now that the multidimensional gravitational constant $\tilde G_{\mathcal{D}}$ and the newtonian gravitational constant $G_N$ are related as
\be{4.19} c^4k_{\mathcal{D}}\; \frac{D-2+\omega d_1}{D-1}=4\pi G_N V_{d_1}\, , \ee
then \rf{4.18} is reduced to usual Poisson equation:
\be{4.20} \triangle \varphi = 4\pi G_N \rho^{(d_0)}\, . \ee
Obviously, in \rf{4.19} and \rf{4.20}, it is assumed that $d_0=3$. \rf{4.19} shows that the relation between gravitational constants $\tilde G_{\mathcal{D}}$ and $G_N$
depends on equation of state in the internal space. If we take the dust-like equation of state, then we obtain \rf{1.19}. However, for the black string equation of state
\rf{4.16} (and $d_0=3$) we obtain
\be{4.21} S_D \tilde G_{\mathcal{D}}=4\pi G_N V_{d_1}\, . \ee
If perfect fluid is confined in three-dimensional sphere of a radius $R$, then \rf{4.20} has the following solution in vacuum: $\varphi = -G_N m /r_3$, where $m
=(4/3)\pi R^3 \rho^{(3)}$.

Therefore, in the case of three-dimensional external space, the black string equations of state $p_0=0$ and $p_1=-(1/2)\varepsilon$ are the only possibility to be at the
same level of agreement with the observations as in general relativity. We also indicate in Conclusion that these equations of state satisfy the necessary (but not
sufficient!) condition for stabilization of the internal space.

\section{Conclusion and discussion}

In our paper, we investigated the multidimensional KK models with compact static spherically symmetric matter sources either in delta-shaped form or in the form of
distributed perfect fluid.
 We usually also supposed that matter is uniformly smeared over the extra dimensions. Some of these models have exact solutions (solitons) and for others we obtained the
 metric coefficients in the weak field limit. The main purpose of our paper was to establish the correspondence between these models and observable data. There are a number
 of well known gravitational experiments in solar system (perihelion shift, deflection of light, time delay of radar echoes, PPN parameters) which can be used to get
 restrictions on parameters of considered models. In our previous paper \cite{EZ3} we have shown that the point-like mass matter source strongly contradicts the experiments
 if the number of spatial dimensions $D>3$. It was a surprise for us because such approach has clear physical interpretation and works very well in general relativity
 \cite{Landau}. Therefore, in the present paper we wanted to clarify the reason of it.

 First, we investigated five-dimensional soliton solutions to single out one which corresponds to point-like mass. We found parameters of this solution and demonstrated
 that $T_{00}$  is the only non-zero component of the energy-momentum tensor. It can be treated as dust-like equation of state in all dimensions. Because in the weak field
 limit the metric coefficients exactly coincide with ones for the point-like mass, this soliton solution contradicts the experiments. Our first thought was that the reason
 for this is the delta-shaped form of the source. To check it we considered the model with finite size static spherically symmetric perfect fluid. For astrophysical
 objects (e.g., Sun) it is usually supposed that the energy density is much bigger than pressure. Therefore, we also assumed that considered perfect fluid has the dust-like
 equation of state in all dimensions. However, here we arrived at exactly the same form of the asymptotic metric coefficients as in the case of point-like mass. Therefore,
 this model also contradicts the observations. Thus, both point-like mass model and perfect fluid with dust-like equation of state in all directions failed with
 experiments. The common for both of these models is the same equations of state in our three dimensions and in the extra dimensions. So, all dimensions are treated at equal
 footing. This is the crucial point. To be in agreement with observations, it is necessary to break the symmetry between the external/our and internal spaces.

 To prove it, we investigated conditions under which the solitonic solutions do not contradict the observations. It can be easily done via the parameterized post-Newtonian
 parameters (PPN) $\gamma$ and $\beta$. We found these parameters with the help of asymptotic expression for the solitonic metric coefficients. Parameter $\beta$ exactly
 coincides with one in general relativity but $\gamma$ is not. Additionally, we also obtained the formulas for perihelion shift, deflection of light and time delay of radar
 echoes in the case of soliton solutions. Using the constraint on $\gamma$ which comes from the Cassini spacecraft experiment, we found that the solitonic parameter $k$
 should be very big: $|k|\geq2.3\times10^4$. Roughly speaking, $|k|\to \infty$. In the case of point-like mass soliton $k=2$, which is clearly contrary to this restriction.

There is one very interesting five-dimensional soliton solution that satisfies the condition $|k|\to \infty$. This is so-called black string. Four-dimensional part of
this metrics exactly coincides with Schwarzschild metrics and the extra dimension is flat (the internal space metric coefficient does not depend on any coordinates).
Obviously, the results of mentioned above gravitational experiments in this model exactly coincide with general relativity. Here, $T_{00}$ and $T_{44}$ are the only
non-zero components of the energy-momentum tensor. Moreover, ${T^0}_0=2{T^4}_4$. It can be treated as dust-like equation of state $p_0=0$ in the external space and very
specific equation of state $p_1=-(1/2)\varepsilon$ in the internal space. For a better understanding of black strings, we have considered a static spherically symmetric
perfect fluid with dust-like equation of state $p_0=0$ in $d_0$-dimensional external space and an arbitrary equation of state $p_1=\omega \varepsilon$ in
$d_1$-dimensional internal space. We took the dust-like equation of state in our space because it is usual condition for astrophysical objects like the Sun. We
demonstrated that the demand of an exact correspondence between this model and general relativity automatically leads to equation of state $p_1=-(1/2)\varepsilon$ for
$d_0=3$. The dimension 3 for the external space is a special one. Only in this case parameter $\omega $ does not depend on $d_1$ and equals to $-1/2$. Therefore, in the
case of three-dimensional external space, the black string equations of state $p_0=0$ and $p_1=-(1/2)\varepsilon$ are the only ones which ensure the same level of
agreement with the observations as in general relativity.

Now, we want to stress an additional and very important feature of the black string equations of state. This feature follows from the conclusions of no-go theorem given
in the appendix. This theorem (case II, which relates to ordinary matter in our Universe) clearly shows that the condition of the internal space stabilization requires
the violation of symmetry (in terms of equations of state) between our three dimensions and the extra dimensions. The need for such a violation is especially seen in the
example of radiation. It is well known that radiation satisfies the equation of state $p=(1/3)\varepsilon$. If we assume equality of all dimensions and allow light to
move around all multidimensional space, then equation of state will be $p=(1/D)\varepsilon$, which apparently contradicts the observations for $D>3$. Therefore,
radiation should not move in the extra dimensions. Exactly this situation we have in case II. If we take $\alpha^{(c)}=1$, then we obtain the usual equation of state for
radiation in our Universe $\alpha_0^{(c)}=4/3 \; \rightarrow \; p_0^{(c)}=(1/3)\varepsilon^{(c)}$ and dust in the internal space: $\alpha_i^{(c)}=1 \; \rightarrow \;
p_i^{(c)}=0$. The latter means that the light does not move in the extra dimensions. Such situation is realized if light is localized on a brane \cite{Rubakov}.

Thus it is clear now why models with the same (e.g., dust-like) equation of state in all directions have failed with experiments. In spite of their clear physical
motivation, they violate the condition of the internal space stabilization. On the contrary, the black strings have the dust-like equation of state $p_0=0$ in our space
and equation of state $p_1=-(1/2)\varepsilon$ in the internal space. This is in full agreement with the stability condition. Additionally, multidimensional matter with
such equations of state satisfies the known experimental data. Therefore, we obtained important restrictions on the equation of state of the multidimensional matter in
directions of extra dimensions for the localized sources of this matter.

Nevertheless, we want to add "a fly in the
ointment". It is connected with the latter equation or, equivalently, with non-zero (negative!) component $T_{44}$ of the energy-momentum tensor. In the black string
papers, it is called the black string tension. In paper by Chodos and Detweiler \cite{CDandMP}, they called it the scalar charge. However, as far as we are aware, the
reason for such negative tension is still not clarified.  What does "squeeze out" the ordinary particles (which form the astrophysical objects) from the extra
dimensions? Simple localization on the brane, as in the case of radiation, is not enough for this because it results in the dust-like equation of state in the extra
dimension. It should be something else. Therefore, our answer about a viable KK model is "yes" (keeping in mind the black string models) if we can give a satisfactory
explanation for the non-zero negative tension.

In our opinion, brane-world models are the most promising alternative to the KK models because they naturally break the symmetry between our three-dimensional Universe
and the extra dimensions. These models require special consideration. We intend to clarify this interesting problem in our forthcoming paper.

\



\section*{ACKNOWLEDGEMENTS}

We want to thank Ignatios Antoniadis and Seifallah Randjbar-Daemi for stimulating discussions. This work was supported in part by the "Cosmomicrophysics"
programme of the Physics and Astronomy Division of the National Academy of Sciences of Ukraine and in part by the Brazilian sponsoring agency CAPES. A. Zh. thanks
Lu\'{i}s Crispino and the Faculdade de F\'{i}sica of the Universidade Federal do Par\'{a} for their hospitality during final preparation of this work.

\appendix*
\section{\label{sec:A}Perfect fluid in multidimensional cosmological models: no-go theorem}
\renewcommand{\theequation}{A\arabic{equation}}
\setcounter{equation}{0}

The no-go theorem \cite{Zhuk} leads to important conclusions in our paper. Therefore, it makes sense to give a brief derivation of it in this appendix.

In conventional cosmology matter fields are taken into account in a phenomenological way as a perfect fluid with equal pressure in all three spatial directions. It
provides homogeneous (if energy density and pressure depend only on time) and isotropic picture of the Universe. In multidimensional case we generalize this approach to
a $m-$component perfect fluid with energy-momentum tensor \cite{perfect fluid}
\be{zh2.26} T^M_N=\sum_{c=1}^m{T^{(c)}}^M_N , \ee
\ba{zh2.27} {T^{(c)}}^M_N &=& -{\rm diag\ } (-\rho ^{(c)},
\underbrace{p_0^{(c)},\ldots ,p_0^{(c)}}_{\mbox{$d_0$ times}%
},\ldots ,\underbrace{p_n^{(c)},\ldots ,p_n^{(c)}}_{%
\mbox{$d_n$ times}}\, )\, .\nn\\
&{}&\ea

In this appendix we accept for convenience the system of units where the speed of light $c\equiv 1$. Thus the energy density $\varepsilon ^{(c)} = \rho^{(c)} c^2 $
coincides with the mass density $\rho^{(c)}$. To get the conditions of the internal space stabilization, we consider the case of dynamical energy density and pressure:
 $\rho ^{(c)} =\rho ^{(c)}(\tau )$ and  $p_i^{(c)}=p_i^{(c)}(\tau ),\quad i=0,\ldots ,n$.

The conservation equations we impose on each component separately:
\be{zh2.28} {T^{(c)}}^M_{N;M} = 0.
\ee
The metrics of spacetime is also taken in the homogeneous form:
\ba{zh2.29}
 g&=&  g^{\, (0)}(x)-\sum_{i=1}^n  e^{2\beta
^i(\tau )}g^{(i)}(y) \\ \nn &\equiv & e^{2\gamma (\tau )}d\tau \otimes d\tau - e^{2\beta ^0(\tau )}g^{(0)}(\vec{x}) - \sum_{i=1}^n e^{2\beta ^i(\tau )}g^{(i)}(y) \, .
\ea
The choice of the function $\gamma (\tau)$ defines different gauges, e.g. the synchronous time gauge $\gamma =0$ or the conformal time gauge $\gamma (\tau)
=\beta^0(\tau)$, etc. In what follows, we use the notations $a \equiv e^{\beta^0}$ and $b_i \equiv e^{\beta^i} \, (i =1,\ldots ,n)$ to describe scale factors of the
external and internal spaces, respectively.

Denoting by a dot differentiation with respect to time $\tau$,
the conservation equations \rf{zh2.28}  for the tensors (\ref{zh2.27}) read
\be{zh2.30}
\dot\rho ^{(c)}+\sum_{i=0}^nd_i\dot \beta ^i\left( \rho
^{(c)}+p_i^{(c)}\right) =0\, .
\ee
If the pressures and energy density are related via equations of state
\be{zh2.31}
p_i^{(c)}=\left( \alpha _i^{(c)}-1\right) \rho ^{(c)},\ \ \ \ \
i=0,\ldots ,n,\quad c=1,\ldots ,m\, ,
\ee
then \rf{zh2.30} has the simple integral
\be{zh2.32}
\rho ^{(c)}(\tau )=A^{(c)}a^{-d_0 \alpha_0^{(c)}}\times
\prod_{i=1}^n b_i^{-d_i\alpha _i^{(c)}}\, ,
\ee
where $A^{(c)}$ is the constant of integration.

To investigate the problem of the stable compactification, it is helpful to use the equivalence between the Einstein equations and the Euler-Lagrange equations for
Lagrangian obtained by dimensional reduction of the action
\be{zh2.33}
S=\frac 1{2\kappa_{\mathcal{D}}
}\int\limits_Md^{\mathcal{D}}x\sqrt{|g|}\left\{ R[g]-2\Lambda_{\mathcal{D}} \right\}
- \int\limits_Md^{\mathcal{D}}x\sqrt{|g|} \rho\, ,
\ee
where $\rho $ is given by \rf{zh2.32} (see \cite{perfect fluid} for details). This equivalence takes place for homogeneous model \rf{zh2.29}.  However, we can generalize
it to the inhomogeneous case allowing inhomogeneous fluctuations $\tilde \beta^i (x) \, (i=1,\ldots ,n)$ over stably compactified background $\beta_0^i = \mbox{const}:
\; \tilde \beta^i = \beta^i - \beta^i_0$. The dimensional reduction (see, e.g., \cite{RZ}-\cite{GSZ}) of the action \rf{zh2.33} results in effective theory in Einstein
frame with the effective potential
\ba{zh2.34}
U_{eff}&=&{\left( \prod_{i=1}^ne^{d_i\tilde \beta ^i}\right) }^{-
2/(D_0-2)}\left[ -\frac 12\sum_{i=1}^n\tilde R_ie^{-2\tilde \beta
^i}\right.\nn \\
&+&\left. \Lambda_{\mathcal{D}} +\kappa _{\mathcal{D}}\sum_{c=1}^m\rho ^{(c)}\right]\, ,
\ea
where $\rho^{(c)}$ is defined by \rf{zh2.32} and $D_0=d_0+1$. The internal space conformal fluctuations $\tilde \beta ^i$ play the role of minimal scalar fields with
potential \rf{zh2.34} in external spacetime with the Einstein frame metrics $\tilde g^{(0)}_{\mu\nu} =\Omega^{-2} g^{(0)}_{\mu\nu}$ where $\Omega^{2}={\left(
\prod_{i=1}^ne^{d_i\tilde \beta ^i}\right) } ^{-2/(D_0-2)}$. Now, we suppose that the external spacetime metrics $\tilde g^{(0)}$  has also the
Friedmann-Robertson-Walker form:
\ba{zh2.35}
\tilde g^{(0)}&=&\Omega ^{-2} g^{(0)}=\tilde g_{\mu \nu
}^{(0)}dx^\mu \otimes dx^\nu \nn \\
&=&e^{2\widehat{ \gamma }}d\widehat{ \tau} \otimes d\widehat{ \tau} - e^{2\widehat{ \beta} ^0(x)}g^{(0)}\, . \ea
It results in the following connection between the external scale factors in the Brans-Dicke frame $a\equiv e^{\beta ^0}$ and in the Einstein frame $\tilde a \equiv
e^{\widehat{ \beta }^0}$:
\be{zh2.36}
a={\left( \prod_{i=1}^ne^{d_i\tilde \beta ^i}\right) }^{-
1/(D_0-2)}\tilde a\, ,
\ee
then, the expression \rf{zh2.32} for $\rho^{(c)}$ can be rewritten in the form
\be{zh2.37}
\kappa_{\mathcal{D}} \rho ^{(c)}=
\kappa_N \rho^{(c)}_{(4)}\prod_{i=1}^n e^{-\xi_i^{(c)}\tilde
\beta^i}\, ,
\ee
where
\be{zh2.38}
\rho^{(c)}_{(4)} = \tilde A^{(c)} \tilde a^{-d_0\alpha_0^{(c)}}\,
, \quad \tilde A^{(c)} = A^{(c)} V_I \prod_{i=1}^n
b_{(0)i}^{d_i(1-\alpha_i^{(c)})}\, ,
\ee
\be{zh2.39}
\xi_i^{(c)} = d_i\left( \alpha _i^{(c)}-\frac{\alpha _0^{(c)}d_0}{%
d_0-1}\right)\,
\ee
and we take into account the relation \rf{4.21} $\kappa_{\mathcal{D}}=\kappa_N V_{D'}$ with the internal space volume $V_{D'} =V_I \prod_{i=1}^n b_{(0)i}^{d_i}$.
$D'=\sum_i^n d_i$ is the total number of the internal dimensions. Prefactor $V_I$ is defined by the geometry of the internal spaces \cite{Zhuk} and for torii $V_I=1$. It can
be easily verified that $\tilde A^{(c)}$ has dimension $\mbox{cm}^{d_0\alpha_0^{(c)}-D_0}$.

Thus, the problem of stabilization of the extra dimensions is reduced now to search of minima of the effective potential $U_{eff}$ with respect to the fluctuations
$\tilde \beta^i$:
\ba{zh2.40}
&{}&\left.\frac{\partial U_{eff}}{\partial \tilde \beta^k}\right|_{\tilde
\beta =0} =0 \Longrightarrow \tilde R_k = -\frac{d_k}{D_0-2}\left[
\sum_{i=1}^n \tilde R_i -2\Lambda_{\mathcal{D}} \right] \nn \\
&+& \kappa_N
\sum_{c=1}^m \rho^{(c)}_{(4)}\left(\xi^{(c)}_k +
\frac{2d_k}{D_0-2}\right) \, , \; k = 1,\ldots ,n\, .
\ea
The left-hand side of this equation is a constant but the right-hand side is a dynamical function because of dynamical behavior of the effective four-dimensional energy
density $\rho^{(c)}_{(4)}$ . Thus, we arrive at the following
{\it no-go theorem}:\\


{\it Multidimensional cosmological Kaluza-Klein models with the perfect fluid as a matter source do not admit stable compactification of the internal spaces with
exception of two special cases:}

\ba{zh2.41}
\mbox{I.} \quad \alpha^{(c)}_0 &=& 0\; ,\quad  \forall \;
\alpha_i^{(c)}
\, \, .\\
&\vphantom{\int}& \nn \\
\label{zh2.42}\mbox{II.}\quad
\xi_i^{(c)}&=&-\frac{2d_i}{d_0-1} \vphantom{\int} \Longrightarrow
\left\{\begin{array}{rcl} \alpha _0^{(c)}&=&\frac
2{d_0}+ \frac{d_0-1}{d_0}\alpha ^{(c)}\, ,\\
&\vphantom{\int}&\\
\alpha _i^{(c)}&=&\alpha ^{(c)}\, .\\\end{array}\right.
\ea
where $i=1,\ldots ,n;\; c=1,\ldots ,m$.

\vspace{0.3cm}

\noindent First case corresponds to vacuum in the external space $\rho^{(c)}_{(4)} = \tilde A^{(c)}=\mbox{const}$ and arbitrary equations of state in the internal
spaces. Some bulk matter can mimic such behavior, e.g., vacuum fluctuations of quantum fields (Casimir effect) \cite{PRD(1997),GKZ}, monopole form fields
\cite{PRD(1997),GMZ} and gas of branes \cite{Kaya}.

In the second case, the energy density in the external space is not a constant but a dynamical function with the following behavior:
\be{zh2.43}
\rho _{(4)}^{(c)}(\tilde a)=\tilde A^{(c)}\frac 1{\tilde
a^{2+(d_0-1)\alpha ^{(c)}}} =\left. \tilde A^{(c)}\frac 1{\tilde
a^{2(1 +\alpha ^{(c)})}}\right|_{d_0=3}\, .
\ee
The corresponding equation of state is:
\be{zh2.44}
p_{(4)}^{(c)} = (1/3)(2\alpha^{(c)}-1)\rho _{(4)}^{(c)}= (\alpha_0^{(c)}-1)\rho _{(4)}^{(c)}\, ,
\ee
where we put $d_0=3$. It can be easily seen from \rf{zh2.37} that in the case of stabilized internal spaces (i.e. $\tilde \beta^i =0$) $\rho _{(4)}^{(c)} = \rho
^{(c)}V_{D'}$. Similar relation takes place for $p_{(4)}^{(c)}$ and  $p_{0}^{(c)}$: $\; p_{(4)}^{(c)} = p_{0}^{(c)}V_{D'}$. Therefore, the second case corresponds to
ordinary matter in our three-dimensional space. For example, in the case of three-dimensional external space, the choice $\alpha ^{(c)}=1/2$ provides dust in our space:
$\alpha_0^{(c)}=1 \; \rightarrow \; p_{(4)}^{(c)}=0$ and equation of state $\alpha_i^{(c)}=1/2 \; \rightarrow \; p_i^{(c)}=-(1/2)\rho^{(c)}$ in the internal spaces,
which are exactly the black string equations of state!

It is worth of noting that the cases I and II are the necessary but not sufficient conditions for stabilization. In \cite{Zhuk} it was shown that stability is ensured by
the matter from the first case with a proper choice of the parameters of models. The matter related to the second case provides the standard evolution of the Universe
and does not spoil the stabilization. As we mentioned above, matter of the first case behaves as a cosmological constant in the external space. There are strong
experimental restrictions on cosmological constant in solar system \cite{SolarCosmConst}. Therefore, there is no need to take into account such perfect fluid for
considered in this paper astrophysical problems. It plays an essential role on cosmological scales.



\begin{thebibliography}{99}

\bibitem{Polchinski}
J. Polchinski, {\it String Theory, Volume 2: Superstring Theory and Beyond}, (Cambridge University Press, Cambridge, 1998).
\bibitem{EZ3}
M. Eingorn and A. Zhuk, Class. Quant. Grav. {\bf 27}, 205014 (2010); (arXiv:gr-qc/1003.5690).
\bibitem{Landau}
L.D. Landau and E.M. Lifshitz, {\it The Classical Theory of Fields, Fourth Edition, Volume 2 (Course of Theoretical Physics Series)}, (Pergamon Press, Oxford, 2000).
\bibitem{EZ1}
M. Eingorn and A. Zhuk, Phys. Rev. D {\bf 80}, 124037 (2009); (arXiv:hep-th/0907.5371).
\bibitem{EZ2}
M. Eingorn and A. Zhuk, Class. Quant. Grav. {\bf 27}, 055002 (2010); (arXiv:gr-qc/0910.3507).
\bibitem{Kramer}
D. Kramer, Acta Phys. Polon. B {\bf 2}, 807 (1970).
\bibitem{soliton}
D.J. Gross and M.J. Perry, Nucl. Phys. B {\bf 226}, 29 (1983).
\bibitem{Davidson}
A. Davidson and D. Owen, Phys. Lett. {\bf 155}, 247 (1985).
\bibitem{Ivashchuk}
V.D. Ivashchuk, {\it Spherically-symmetric solutions with a chain of n internal Ricci-flat spaces} (2010); (arXiv:gr-qc/1006.4605).
\bibitem{FIM}
S.B. Fadeev, V.D. Ivashchuk and V.N. Melnikov, Phys. Lett. A {\bf 161}, 98 (1991); (arXiv:gr-qc/1006.5147).
\bibitem{Leon}
J.P. de Leon, Grav. Cosmol. {\bf 15}, 345 (2009); (arXiv:gr-qc/0905.2010); J.P. de Leon, {\it Schwarzschild-like exteriors for stars in Kaluza-Klein gravity} (2010);
(arXiv:gr-qc/1003.3151).
\bibitem{Will}
C.M. Will, {\it Theory and Experiment in Gravitational Physics}, (Cambridge University Press, Cambridge, 2000).
\bibitem{Straumann}
N. Straumann, {\it General Relatuvuty and Relativistic Astrophysics}, (Springer-Verlag, Berlin, Heidelberg, 1984).
\bibitem{Will2}
C.M. Will, {\it Was Einstein Right? Testing Relativity at the Century}. In {\it 100 Years of Relativity: Spacetime Structure -- Einstein and Beyond, ed. Abhay Ashtekar,
page 205}. (World Scientific, Singapore, 2005); (arXiv:gr-qc/0504086).
\bibitem{JKh}
Bh. Jain and J. Khoury, {\it Cosmological Tests of Gravity} (2010); (arXiv:astro-ph/1004.3294).
\bibitem{Bertotti}
B. Bertotti, L. Iess and P. Tortora, Nature {\bf 425}, 374 (2003).
\bibitem{Harmark}
T. Harmark and N.A. Obers, JHEP {\bf 0405}, 043 (2004); (arXiv:hep-th/0403103).
\bibitem{Traschen}
D. Kastor and J. Traschen, JHEP {\bf 0609}, 022 (2006); (arXiv:hep-th/0607051).
\bibitem{Zhuk}
A. Zhuk, {\it Conventional cosmology from multidimensional models \/}. In {\it Proceedings of the 14th International Seminar on High Energy Physics "QUARKS-2006" in St.
Petersburg (May 19-25 2006), vol. 2, page 264}. (INR press, 2007); (arXiv:hep-th/0609126).
\bibitem{Rubakov}
V.A. Rubakov, Phys. Usp. {\bf 44}, 871-893 (2001); Usp. Fiz. Nauk {\bf 171}, 913-938 (2001); (arXiv:hep-ph/0104152).
\bibitem{CDandMP}
A. Chodos and S. Detweiler, Gen. Rel. Grav. {\bf 14}, 879 (1982).
\bibitem{perfect fluid}
V.D. Ivashchuk and V.N. Melnikov, Int. Journ. Mod. Phys. D {\bf 3}, 795 (1994); (arXiv:gr-qc/9403064); V.D.~Ivashchuk and V.N. Melnikov, Class. Quant. Grav. {\bf 12},
809 (1995); (arXiv:gr-qc/9407028); U. Kasper and A. Zhuk, General Relativity and Gravitation {\bf 28}, 1269-1292 (1996); A. Zhuk, Classical and Quantum Gravity {\bf 13},
2163-2178 (1996).
\bibitem{RZ}
M. Rainer and A. Zhuk, Phys. Rev. D {\bf 54}, 6186-6192, (1996); (arXiv:gr-qc/9608020).
\bibitem{PRD(1997)}
U. G\"unther and A. Zhuk, Phys. Rev. D {\bf 56}, 6391-6402, (1997); (arXiv:gr-qc/9706050).
\bibitem{GSZ}
U. G\"unther, A. Starobinsky and A. Zhuk, Phys. Rev. D {\bf 69}, 044003 (2004); (arXiv:hep-ph/0306191).
\bibitem{GKZ}
U. G\"unther, S. Kriskiv and A. Zhuk, Grav. Cosmol. {\bf 4}, 1 (1998); (arXiv:gr-qc/9801013).
\bibitem{GMZ}
U. G\"unther, P. Moniz and A. Zhuk, Phys. Rev. D {\bf 68}, 044010 (2003); (arXiv:hep-th/0303023).
\bibitem{Kaya}
A. Kaya, JCAP {\bf 0408}, 014 (2004); (arXiv:hep-th/0405099).
\bibitem{SolarCosmConst}
P. Jetzer and M. Sereno, {\it Solar system tests of the cosmological constant} (arXiv:astro-ph/0711.3723); L. Iorio, {\it Solar System motions and the cosmological
constant: a new approach} (arXiv:gr-qc/0710.2610).




\end{thebibliography}
\end{document}